\RequirePackage[2020-02-02]{latexrelease}
\documentclass[aps,prd,groupedaddress,graphicx,nofootinbib]{revtex4}
\usepackage{amsmath,amssymb,graphics,graphicx,color,epsf}
\usepackage{subfigure}
\usepackage[scanall]{psfrag}  
\usepackage{tikz}

\begin{document}

\title{A note on the (non-)conservation of curvature perturbation}

\author{Chia-Min Lin}

\affiliation{Fundamental General Education Center, National Chin-Yi University of Technology, Taichung 41170, Taiwan}



\begin{abstract}

In this note, we compare two different definitions for the cosmological perturbation $\zeta$ which is conserved on large scales, and study their non-conservation on small scales. We derive an equation for the time evolution of the curvature perturbation on a uniform density slice through a calculation solely in longitudinal (conformal-Newtonian) gauge. The result is concise and compatible with that obtained via local conservation of energy-momentum tensor.

\end{abstract}
\maketitle
\large
\baselineskip 18pt
\section{Introduction}

Our universe appears to be homogeneous and isotropic on large enough scales in accordance with cosmological principle and is described by an expanding Friedmann metric. On smaller scales, there are inhomogeneities that can be studied by using perturbation theory. These primordial perturbations generate isotropy of the cosmic microwave background (CMB) and are the seeds of subsequent structure formation which eventually lead to galaxies and stars. But inhomogeneity of what? Naively we can say it is the inhomogeneity of energy density, but in cosmological perturbation theory, one has the freedom to choose a spatial hypersurface (or slice) where the energy density is constant (or uniform) and there is no density perturbation at all. Instead, the hypersurface thus chosen may have intrinsic curvature. There is a give and take between primordial density perturbation and primordial curvature perturbation. They can be transformed into each other. Therefore cosmological perturbation theory is complicated by the issue of coordinate (or gauge) transformation and there are different theoretical representations for the same physics. The question about which gauge is better has an answer up to personal taste. One popular approach is to choose a longitudinal (conformal-Newtonian) gauge and study the time evolution of a gauge invariant quantity $\Phi$ which can be regarded as a generalized Newtonian potential. Another approach is to consider a quantity $\zeta$ which is conserved on large scales, but not so conserved on small scales. The purpose of this note is to consider two definitions of $\zeta$ (we call them $\zeta_1$ and $\zeta_2$) and compare their non-conservation on small scales. 

The note is organized as follows.
We introduce cosmological perturbation theory in Section \ref{sec1}.
We review some relevant equations which will be needed subsequently in order to work in longitudinal gauge in Section \ref{sec2}. We present the time evolution and matching conditions for $\zeta_1$ in order to compare with $\zeta_2$ in Section \ref{sec3}. We solve for the time evolution of $\zeta_2$ and obtain a concise equation in Section \ref{sec4}. The calculations are done solely in the framework of a longitudinal (conformal-Newtonian) gauge. The results can be compared with that of other approaches. In Section \ref{sec5}, we present our conclusions.

\section{cosmological perturbation}
\label{sec1}
We consider a spatially flat Friedmann metric $^{(0)}g_{\mu\nu}$ with first-order scalar perturbations $\delta g_{\mu\nu}$. The line element is given by
\begin{equation}
d s^2= a^2 \left[ (1+2A)d\eta^2 + 2B_{,i}d\eta dx^i -(1-2\psi \delta_{ij} -2E_{,ij})dx^i dx^j  \right],
\end{equation}
With the coordinate transformation
\begin{equation}
x^\alpha \rightarrow x^\alpha + \xi^\alpha,
\end{equation}
the variation of the metric perturbation $\delta g_{\mu\nu}$ (up to first-order) is nothing but the Lie derivative of the background metric $^{(0)}g_{\mu\nu}$ with respect to the vector $\xi^\alpha$, namely $\delta g_{\mu\nu} \rightarrow \delta g_{\mu\nu}+\mathsterling_\xi ^{(0)}g_{\mu\nu}$, where
\begin{equation}
\mathsterling_\xi ^{(0)}g_{\mu\nu}=\xi^\lambda  {^{(0)}g_{\mu\nu,\lambda}}+^{(0)}g_{\lambda \nu}\xi^\lambda_{,\mu}+ ^{(0)}g_{\mu\lambda} \xi^\lambda_{,\nu}.
\end{equation}
If we define $\xi^\alpha \equiv (\xi^0, \xi^i)$ where $\xi^i$ is decomposed through Helmholtz's theorem as
\begin{equation}
\xi^i=\xi^i_{\perp}+\xi^{,i},
\end{equation}
with $\xi^i_{\perp,i}=0$, the gauge transformation of the metric perturbation is given by
\begin{equation}
A \rightarrow A-\frac{1}{a}(a\xi^0)^\prime, \;\;\;  B \rightarrow B+\xi^\prime -\xi^0, \;\;\; \psi \rightarrow \psi+\frac{a^\prime}{a} \xi^0, \;\;\; E \rightarrow E+ \xi,
\label{gauge}
\end{equation}
where a dash indicates differentiation with respect to $\eta$.

For the unperturbed background, we have the continuity equation
\begin{equation}
\epsilon^\prime_0=-3\mathcal{H}(\epsilon_0+p_0).
\label{cl}
\end{equation}
and the Friedmann equation 
\begin{equation}
\mathcal{H}^2=\frac{8\pi G}{3}a^2 \epsilon_0,
\label{e2}
\end{equation}
where $\mathcal{H}\equiv a^\prime/a$ is the Hubble parameter\footnote{This is different from $H \equiv \frac{(da/dt)}{a}$ where $dt=ad\eta$.}, $\epsilon_0$ is the unperturbed energy density, and $p_0$ is the unperturbed pressure.
These equations will be used in the following sections.
\section{Longitudinal (conformal-Newtonian) gauge}
\label{sec2}
From Eq.~(\ref{gauge}), it can be seen that gauge transformation can be used to set $B=E=0$, and the gauge freedom is used up\footnote{Two scalar degrees of freedom $\xi^0$ and $\xi$ are used to cancel two scalar degrees of freedom $B$ and $E$.}. In this particular gauge (longitudinal or conformal-Newtonian gauge), we call $A=\Phi$ and $\psi=\Psi$\footnote{They are gauge-invariant quantities.}. 
The line element is then simplified to
\begin{equation}
ds^2=a^2 \left[ (1+2\Psi)d \eta^2-(1-2\Phi)\delta_{ij}dx^i dx^j \right]
\end{equation}
The linearized Einstein equations are
\begin{equation}
\overline{\delta G}^\alpha_\beta=8\pi G \overline{\delta T}^\alpha_\beta,
\end{equation}
where the overline symbol denotes perturbations calculated in this particular gauge and can be defined in a gauge-invariant way. This gives \cite{Mukhanov:2005sc, Mukhanov:1990me}
\begin{equation}
\Delta \Psi - 3 \mathcal{H}(\Psi^\prime+\mathcal{H}\Phi)=4 \pi G a^2 \overline{\delta T}^0_0,
\label{ee11}
\end{equation}
\begin{equation}   
(\Psi^\prime +\mathcal{H}\Phi)_{,i}=4\pi G a^2 \overline{\delta T}^0_i,
\label{e12}
\end{equation}
\begin{equation}
\left[ \Psi^{\prime\prime} +\mathcal{H}(2\Psi +\Phi)^\prime +(2\mathcal{H}+\mathcal{H}^2)\Phi + \frac{1}{2}\Delta (\Phi-\Psi) \right] \delta_{ij}-\frac{1}{2}(\Phi-\Psi)_{,ij}=-4 \pi G a^2 \overline{\delta T}^i_j.
\label{e13}
\end{equation}
Here the symbol $\Delta$ denotes the Laplacian for the comoving spatial coordinates $x^i \equiv  \mathbf{x}$. It corresponds to $-k^2$ in momentum space, where $k \equiv |\mathbf{k}|$ is the comoving wave number for a mode $\propto \exp(i \mathbf{kx}) $.
The energy-momentum tensor is written as 
\begin{equation}
\overline{\delta T}^0_0=\overline{\delta \epsilon}, \;\;\; \overline{\delta T}^0_i=\frac{1}{a}(\epsilon_0+p_0)(\overline{\delta u}_{\| i}), \;\;\; \overline{\delta T}^i_j=-\overline{\delta p}\delta ^i_j.
\label{eq14}
\end{equation}
Here we define $\delta u_{\|i}\equiv \delta u_{\|,i}$ for a scalar function $\delta u_\|$ and \footnote{We ignore another component $\delta u_{\perp i}$ with the property $(\delta u_{\perp i})^{,i}=0$ because it does not affect scalar perturbation.} 
\begin{equation}
\overline{\delta u}_{\|i}=\delta u_{\|i}-a(B-E^\prime)_{,i}.
\label{eqgt}
\end{equation}
From Eq.~(\ref{eq14}), $\overline{\delta T}^i_j=0$ for $i \neq j$ therefore $\Psi=\Phi$ from Eq.~(\ref{e13}). By using these results, Eqs.~(\ref{ee11}), (\ref{e12}), and (\ref{e13}) are simplified to
\begin{equation}
\Delta \Phi-3\mathcal{H}(\Phi^\prime+\mathcal{H}\Phi)=4\pi G a^2 \overline{\delta \epsilon},
\label{de}
\end{equation}
\begin{equation}
(\Phi^\prime+\mathcal{H}\Phi)_{,i}=4\pi G a (\epsilon_0+ p_0)\overline{\delta u}_{\| i},
\label{eq9}
\end{equation}
\begin{equation}
\Phi^{\prime\prime}+3 \mathcal{H}\Phi^\prime +(2 \mathcal{H}^\prime + \mathcal{H}^2)\Phi=4\pi G a^2 \overline{\delta p}.
\label{e11}
\end{equation}
The pressure is a function of energy density $\epsilon$ and entropy $S$, hence the perturbation is
\begin{equation}
\overline{\delta p}=c^2_s \overline{\delta \epsilon}+\tau \delta S,
\label{state}
\end{equation}
where $c^2_s \equiv (\partial p/\partial \epsilon)_S$ and $\tau \equiv (\partial p / \partial S)_\epsilon$. From Eqs.~(\ref{de}) and (\ref{e11}), we have
\begin{equation}
\Phi^{\prime\prime}+3(1+c^2_s)\mathcal{H}\Phi^\prime -c^2_s \Delta \Phi +(2\mathcal{H}^\prime +(1+3c_s^2)\mathcal{H}^2)\Phi=4\pi G a^2 \tau \delta S.
\label{main}
\end{equation}
We will consider adiabatic perturbations where $\delta S=0$ in the following discussion. The first derivative term in the above equation can be eliminated if we define
\begin{equation}
u \equiv \frac{\Phi}{(\epsilon_0+p_0)^{1/2}}
\label{u}
\end{equation}
and
\begin{equation}
\theta \equiv \frac{1}{a}\left( 1+\frac{p_0}{\epsilon_0} \right)^{-1/2}.
\label{theta}
\end{equation}
By using $u$ and $\theta$, Eq.~(\ref{main}) becomes
\begin{equation}
u^{\prime\prime}-c_s^2 \Delta u - \frac{\theta^{\prime\prime}}{\theta}u=0,
\label{eq2}
\end{equation}
which can be further rearranged into
\begin{equation}
\left[ \theta^2 \left( \frac{u}{\theta} \right)^\prime \right]^\prime = c^2_s \theta^2 \Delta \left( \frac{u}{\theta} \right).
\label{Del}
\end{equation}
Let us define the quantity $\zeta_1$ as
\begin{equation}
\zeta_1 \equiv \frac{2}{3}\frac{\mathcal{H}^{-1}\Phi^\prime+\Phi}{1+w}+\Phi.
\label{zeta1}
\end{equation}
This definition is used in modern textbooks and reviews such as \cite{Mukhanov:2005sc, Lyth:2009zz, Brandenberger:1994ce, Mukhanov:1990me, Durrer:2004fx}.
By using Eqs.~(\ref{u}) and (\ref{theta}), we can obtain
\begin{equation}
\zeta_1 = \frac{2}{3}\left( \frac{8\pi G}{3} \right)^{-1/2}\theta^2 \left( \frac{u}{\theta} \right)^\prime.
\label{z12}
\end{equation}
From Eq.~(\ref{Del}), the time derivative of $\zeta$ is given by
\begin{equation}
\zeta_1^\prime = \frac{2}{3}\left( \frac{8\pi G}{3} \right)^{-1/2}c_s^2 \theta^2 \Delta \left( \frac{u}{\theta} \right).
\label{z12p}
\end{equation}
For large scales (or the comoving wavenumber $k\rightarrow 0$ or $\Delta \Phi \rightarrow 0$), $\zeta^\prime=0$ and it is a useful conserved quantity.
We would like to study the effect of non-zero $k$ in the following.

\section{(violation of) the conservation of $\zeta_1$}
\label{sec3}
For a mode with wavenumber $k$, Eq.~(\ref{eq2}) can be rewritten as
\begin{equation}
u_k(\eta)=C_1 \theta + C_2 \theta \int \frac{d\eta}{\theta^2}-k^2 \theta \int^\eta \left( \int^{\tilde{\eta}} c_s^2 \theta u_k d\bar{\eta} \right)\frac{1}{\theta^2(\tilde{\eta})}d\tilde{\eta}.
\end{equation}

From Eq.~(\ref{z12}), we have\footnote{This appears as an exercise in \cite{Mukhanov:2005sc}.}
\begin{equation}
\zeta_1=\frac{2}{3}\left( \frac{8\pi G}{3} \right)^{-1/2} C_2 - \frac{2}{3}\left( \frac{8\pi G}{3} \right)^{-1/2} k^2 \int^\eta c_s^2 \theta u_k d\bar{\eta}.
\label{vio}
\end{equation}
The first term is just a constant and the second term depends on $\eta$ and explicitly shows the violation of the otherwise conserved quantity $\zeta_1$. 
It can also be obtained by integrating Eq.~(\ref{z12p}).

If the pressure $p(\epsilon)$ is discontinuous on the hypersurface $\Sigma$, matching conditions \cite{Mukhanov:2005sc, Deruelle:1995kd} can be developed by integrating Eq.~(\ref{Del}) near $\Sigma$ as
\begin{equation}
\left[ \theta^2 \left( \frac{u}{\theta} \right)^\prime \right]_{\pm}=\int^{\Sigma+0}_{\Sigma-0} c^2_s \theta^2 \Delta \left( \frac{u}{\theta} \right) d \eta,
\end{equation}
where $[X]_{\pm} \equiv X_+ - X_-$.
By using the relation (which is derived in the appendix)
\begin{equation}
c_s^2 \theta^2 =\frac{\epsilon_0}{3 a^2 \mathcal{H}}\left( \frac{1}{\epsilon_0+p_0} \right)^\prime - \frac{\epsilon_0}{a^2 (\epsilon_0+p_0)},
\label{re}
\end{equation}
and continuity of $a$, $\epsilon$, and $u/\theta$ one obtains the matching conditions
\begin{equation}
[\Phi]_{\pm}=0,     \;\;\;   \left[ \zeta_1-\frac{2}{9\mathcal{H}^2}\frac{\Delta \Phi}{1+w},    \right]_{\pm}=0
\label{m2}
\end{equation}
where $w=p_0/\epsilon_0$.
Only for long-wavelength perturbations when $\Delta \Phi$ can be neglected, we have
\begin{equation}
[\zeta_1]_{\pm}=0.
\end{equation} 

\section{(violation of) the conservation of $\zeta_2$}
\label{sec4}
The curvature perturbation on a uniform density slice is defined as\footnote{This quantity is originated in \cite{Bardeen:1983qw} defined in the uniform expansion gauge. There could be a minus sign difference in the definitions. A comparison of $\zeta_1$ and $\zeta_2$ is discussed in \cite{Martin:1997zd} where they are called $\zeta$ and $\zeta_{BST}$.}
\begin{equation}
\zeta_2  \equiv  \mathcal{H} \frac{\delta \epsilon}{\epsilon_0^\prime}+ \psi,
\end{equation}
which can be calculated in any gauge due to its gauge invariance. Note that $\delta \epsilon$ is the energy perturbation in any gauge. The relation between $\delta \epsilon$ and $\overline{\delta \epsilon}$ is $\overline{\delta \epsilon}=\delta \epsilon-\epsilon^\prime_0(B-E^\prime)$. We have $\overline{\delta \epsilon}=\delta \epsilon$ in the longitudinal gauge where $B=E=0$. If one chooses a gauge where $\delta \epsilon=0$ (uniform density slice), $\zeta_2$ is given by the $\psi$ (curvature perturbation\footnote{It is called curvature perturbation because $\psi$ determines the intrinsic spatial curvature on hypersurfaces of constant $\eta$.}) 
at this gauge hence it is called curvature perturbation in uniform density slice. In particular, we can calculate $\zeta_2$ in the longitudinal (conformal-Newtonian) gauge as
\begin{equation}
\zeta_2= \mathcal{H} \frac{\overline{\delta \epsilon}}{\epsilon_0^\prime}+ \Phi.
\end{equation}
By using Eqs.~(\ref{cl}), (\ref{e2}), and (\ref{de}), we obtain\footnote{This appears as the definition of $\zeta$ in  \cite{Brandenberger:1983tg, Brandenberger:1984cz}.}
\begin{equation}
\zeta_2 = \frac{2}{3}\frac{\mathcal{H}^{-1}\Phi^\prime+\Phi}{1+w}+\Phi-\frac{2}{9\mathcal{H}^2}\frac{\Delta \Phi}{1+w}=\zeta_1-\frac{2}{9\mathcal{H}^2}\frac{\Delta \Phi}{1+w}.
\end{equation}
We find that Eq.~(\ref{m2}) is immediately simplified to
\begin{equation}
[\Phi]_{\pm}=0,     \;\;\;   \left[ \zeta_2   \right]_{\pm}=0.
\label{m4}
\end{equation}
The condition for long-wavelength perturbations is immaterial for $\zeta_2$. How about the violation of conservation?
By using Eqs.~(\ref{u}), (\ref{theta}) and (\ref{re}), we write the integrand of the second term in Eq.~(\ref{vio}) as
\begin{equation}
c^2_s \theta^2 \Delta \left( \frac{u}{\theta} \right)=\frac{\epsilon_0}{3a^2 \mathcal{H}}\left( \frac{1}{\epsilon_0+p_0} \right)^\prime \frac{a \Delta \Phi}{\epsilon_0^{1/2}}-\frac{\epsilon_0}{a^2 (\epsilon_0+p_0)} \frac{a \Delta \Phi}{\epsilon_0^{1/2}}  \equiv \frac{X}{3 \mathcal{H}}\left( \frac{1}{\epsilon_0+p_0} \right)^\prime -X\left( \frac{1}{\epsilon_0+p_0} \right),
\label{e29}
\end{equation}
where we have defined
\begin{equation}
X \equiv \frac{\Delta \Phi \epsilon_0^{1/2}}{a}=\sqrt{\frac{3}{8 \pi G}}\frac{\Delta \Phi \mathcal{H}}{a^2}
\label{xd}
\end{equation}
to simplify the calculation.
The equality is obtained by using Eq.~(\ref{e2}).

Let us calculate 
\begin{equation}
\zeta_2^\prime=\zeta_1^\prime -\left[ \frac{2}{9 \mathcal{H}^2}\frac{\Delta \Phi}{1+w} \right]^\prime=\frac{2}{3}\left( \frac{8 \pi G}{3} \right)^{-1/2}c_s^2 \theta^2 \Delta \left( \frac{u}{\theta} \right)-\left[ \frac{2}{9 \mathcal{H}^2}\frac{\Delta \Phi}{1+w} \right]^\prime.
\label{z2p}
\end{equation}
We can use the equality (obtained from Eq.~(\ref{e2}))
\begin{equation}
\frac{2}{9\mathcal{H}^2}\frac{\Delta \Phi}{1+w}=  \frac{2}{3} \left( \frac{8 \pi G}{3} \right)^{-1/2}  \frac{\epsilon_0^{1/2} \Delta \Phi}{3a \mathcal{H}}\left( \frac{1}{\epsilon_0+p_0} \right) \equiv \frac{2}{3} \left( \frac{8 \pi G}{3} \right)^{-1/2}  \frac{X}{3 \mathcal{H}}\left( \frac{1}{\epsilon_0+p_0} \right),
\end{equation}
to obtain
\begin{equation}
\left[ \frac{2}{9 \mathcal{H}^2}\frac{\Delta \Phi}{1+w} \right]^\prime=\frac{2}{3} \left( \frac{8 \pi G}{3} \right)^{-1/2} \left\{ \left( \frac{X}{3 \mathcal{H}}\right)^\prime \left( \frac{1}{\epsilon_0+p_0} \right)+  \left( \frac{X}{3 \mathcal{H}}\right) \left( \frac{1}{\epsilon_0+p_0} \right)^\prime   \right\},
\label{e33}
\end{equation}
where direct calculation shows
\begin{equation}
\left( \frac{X}{3 \mathcal{H}} \right)^\prime=X \left( \frac{\Delta \Phi^\prime}{3 \Delta \Phi \mathcal{H}}-\frac{2}{3} \right).
\end{equation}
By substituting Eqs.~(\ref{e29}) and (\ref{e33}) into Eq.~(\ref{z2p}), we obtain
\begin{equation}
\zeta_2^\prime = \frac{2}{3} \left( \frac{8 \pi G}{3} \right)^{-1/2} \frac{X}{3\mathcal{H}\Delta \Phi}(\Delta \Phi^\prime+\mathcal{H} \Delta \Phi)
\end{equation}
Finally, by using Eqs.~(\ref{eq9}) and (\ref{xd}), we obtain
\begin{equation}
\zeta_2^\prime = \frac{\overline{\delta T}^{0,i}_i}{3(\epsilon_0+p_0)}   = \frac{1}{3a}\Delta \overline{\delta u}_{\|}= \frac{1}{3a}\Delta (\delta u_\|-a(B-E^\prime)),
\label{result}
\end{equation}
where the third equality is from Eq.~(\ref{eqgt}).
This is the main result of this note\footnote{There is a similar expression for the time derivative of $\zeta$ in \cite{Wands:2000dp} obtained via a different approach using only the local conservation of energy-momentum tensor without assumption of Einstein gravity.}. 
This concise equation simply shows that in the case of adiabatic perturbations, $\zeta_2$ is conserved whenever we can neglect the right-hand side of the equation. The last equality allows us to calculate $\zeta^\prime_2$ in any gauge. For example, in comoving gauge where $\delta u_\|=B=0$, we have $\zeta_2^\prime=\Delta E^\prime /3$. If we define\footnote{Velocity divergence is considered for example in \cite{Lesgourgues:2013qba} where it is called $\theta$.} a velocity divergence $\Theta$ as $ \delta T^{0,i}_i \equiv(\epsilon_0+p_0)\Theta $, Eq.~(\ref{result}) can be written more succinctly as 
\begin{equation}
\zeta_2^\prime = \frac{\overline{\Theta}}{3}.
\end{equation}
This shows the time evolution of $\zeta_2$ is governed by the velocity divergence in the longitudinal (conformal-Newtonian) gauge.

The equation can be applied to various models. For example, consider a single-field inflation model with a scalar field as the inflaton field $\phi$. The action is given by
\begin{equation}
S=\int \left( \frac{1}{2} g^{\mu\nu} \phi_\mu \phi_\nu -V  \right)\sqrt{-g}d^4 x.
\end{equation}
The scalar field is a quantum field with quantum fluctuation which results in a small perturbation $\delta\phi$. 
The corresponding perturbation of the relevant component of the energy-momentum tensor is 
\begin{equation}
\overline{\delta T}^0_i=\frac{1}{a^2}(\phi^\prime_0 \overline{\delta \phi})_{,i}.
\end{equation}
By using Eq.~(\ref{result}), this immediately gives the equation of motion for $\zeta_2$ as
\begin{equation}
\zeta_2^\prime = \frac{\phi^\prime_0 \Delta \overline{\delta \phi}}{3 a^2 (\epsilon_0 + p_0)}.
\label{sfi}
\end{equation}
On large scales, the Laplacian for $\overline{\delta \phi}$ is small and $\zeta_2$ is a conserved quantity, in this case, Eq.~(\ref{sfi}) would be useful to let us estimate how large the scale needs to be in order to ignore the evolution of $\zeta_2$. On the other hand, it is also useful when we consider the case of small-scale fluctuations. It may also be useful in the study of reheating after inflation.

As another example, let us consider $k$-inflation \cite{Armendariz-Picon:1999hyi} with the action
\begin{equation}
S=\int p(X,\phi)\sqrt{-g}d^4 X,
\end{equation}
where $X=(1/2)g^{\mu\nu} \phi_\mu \phi_\nu$ and $\epsilon=2X p_{,X}-p$.
Similar to the previous case, consider the perturbation of the inflaton field $\phi$ and the corresponding perturbation of the energy-momentum tensor,  we would have
\begin{equation}
\overline{\delta T}^0_i=(\epsilon +p)\left( \frac{\overline{\delta \phi}}{\phi_0^\prime} \right)_{,i}.
\end{equation}
Therefore from Eq.~(\ref{sfi}),
\begin{equation}
\zeta_2^\prime = \frac{ \Delta \overline{\delta \phi}}{3 a^2 \phi_0^\prime}.
\end{equation}
By applying this result, we can clearly know under what condition the time evolution of $\zeta_2$ can be ignored.
\section{conclusion}
\label{sec5}
In this note, we compare two definitions of $\zeta_1$ and $\zeta_2$. We show that the matching condition for $\zeta_2$ is simpler than that of $\zeta_1$. In particular, we derive the time evolution equation for $\zeta_2$ solely in the framework of the longitudinal (conformal-Newtonian) gauge. The result is very concise and can be compared with that obtained through a different approach in the literature. We present two categories of inflation models as examples, but the application of this result is very broad, especially when one is working on their model in the longitudinal (conformal-Newtonian) gauge.

Of course, if we are only interested in large scales where the comoving wave number $k \rightarrow 0$, then $\zeta_1=\zeta_2$ and we could just call them $\zeta$ and it is conserved (for adiabatic perturbations). However, when discussing situations where $k \neq 0$, it is not always clear which definition is used in the literatures and it is good to know when we could neglect those small-scale corrections. For $\zeta_1$ the condition is whether $\Delta \Phi$ can be neglected. On the hand for $\zeta_2$, the condition is whether $\Delta \overline{\delta u}_{\|}/3a$ can be neglected. We believe this note could help to unify the ideas from different approaches.

\appendix
\section{a useful relation}
We derive Eq.~(\ref{re}) here.
Let us start by calculating
\begin{eqnarray}
\left( \frac{1}{\epsilon_0+p_0} \right)^\prime \frac{\epsilon_0}{3a^2 \mathcal{H}}&=&-\frac{\epsilon_0^\prime + p_0^\prime}{(\epsilon_0+p_0)^2}\frac{\epsilon_0}{3a^2 \mathcal{H}}\\
&=&-\frac{\epsilon_0^\prime \left( 1+ \frac{p_0^\prime}{\epsilon_0^\prime} \right)}{(\epsilon_0+p_0)^2}\frac{\epsilon_0}{3a^2 \mathcal{H}}\\
&=&\frac{3\mathcal{H}(\epsilon_0+p_0)\left( 1+\frac{p_0^\prime}{\epsilon_0^\prime}\right)\epsilon_0}{3\mathcal{H}(\epsilon_0+p_0)^2a^2}\\
&=&\frac{\epsilon_0}{a^2 (\epsilon_0+p_0)}\left(\frac{p_0^\prime}{\epsilon_0^\prime}  \right)+\frac{\epsilon_0}{a^2(\epsilon_0+p_0)}\\
&=&c_s^2 \theta^2+\frac{\epsilon_0}{a^2(\epsilon_0+p_0)}.
\end{eqnarray}
Here in the third equality, we have used Eq.~(\ref{cl}). In the last equality, we have used $c_s \equiv (\partial p_0 /\partial \epsilon_0)=p_0^\prime/\epsilon_0^\prime$ and the definition of $\theta$ from Eqs.~(\ref{state}) and (\ref{theta}). Therefore
\begin{equation}
c^2_s \theta^2=\frac{\epsilon_0}{3a^2 \mathcal{H}}\left( \frac{1}{\epsilon_0+p_0} \right)^\prime-\frac{\epsilon_0}{a^2(\epsilon_0+p_0)}.
\end{equation}

\acknowledgments
This work is supported by the National Science and Technology Council (NSTC) of Taiwan under Grant No. NSTC 111-2112-M-167-002.

\end{document}